# Sound and motion in the De Generatione Sonorum, a treatise by Robert Grosseteste

**Amelia Carolina Sparavigna**
Department of Applied Science and Technology, Politecnico di Torino,
Corso Duca degli Abruzzi 24, Torino, Italy

*Here I am proposing a discussion of some parts of the De Generatione Sonorum, one of the short treatises written by Robert Grosseteste. The main subject of this treatise is the phonetics. However, sound and oscillations are shortly described. Since Grosseteste discussed a phonetics based on the analogy with motions and their combinations, we can find in this text some hints on how a medieval scholar approached the description of motions.*

Robert Grosseteste, an English scientist and philosopher of the Middle Age who was Bishop of Lincoln from 1235 AD until his death on 9 October 1253, wrote several short scientific treatises. Recently we have discussed the De Iride, on optics [1]. Here we propone the discussion of another short paper on sound and phonetics, the "De generatione sonorum". In fact, the sound is shortly discussed at the beginning of the treatise, which quite soon departs from the sound to arrive into the main topic, the phonetics, which is the creation of sounds by the mouth to have syllables and words. The treatise is interesting, because, besides some hints on the physics of sound and oscillations, it contains some descriptions of motions in the discussion of vowels and consonants. That is, Grosseteste uses an analogy with motions and their combinations, rectilinear and circular, and vertical and transversal, in describing how the breath and the organs of speech are creating the voice.

In this medieval treatise we find a list of motions, subdivided in seven types. These are not the seven types of motion (up/down, right/left, forward/back, rotary), proposed by Philo of Alexandria in his treatise On Creation [2]. The Grosseteste's motions are quite complex: it is possible that during the renaissance of sciences in the 12th century, stimulated by the translation of ancient Greek and Arabic works, some scholars started developing a language suitable to describe what happens in mechanics and technology. In fact, the motions given by Grosseteste can be applied to approximately describe how parts of some machines can move. This treatise on sound and voice is therefore remarkable because we can find in it how a medieval scholar approached the description of motions.

In the following, some parts of the Latin text are given in the MS UI Gothic characters, followed by a translation The author is giving a complete translation of the Grosseteste's treatise in Ref.[3].

The treatise starts with the sound generation. Grosseteste tells:

> Cum sonativum percutitur violenter, partes ipsius sonativi egrediuntur a situ naturali, quas natura sonativi reinclinat ad situm naturalem et fortitudine inclinationis huiusmodi partes egressae a situ naturali sibi in toto redeuntes ad situm (sibi) naturalem etiam ultra progrediuntur ad situm sibi non naturalem; et inclinatio naturalis de novo via conversa reinclinat ad situm naturalem et sic fit tremor subtilis in extimis partibus sonativi. Et hic tremor manifestus est in tactu et visu.

*When a medium able to sound is struck violently, some elements of this medium are moving out from their natural positions, to which they come back constrained by the nature of the medium, and, in such a manner, because of the strength of the attraction, by which the moving parts are affected, they completely return at their natural positions, and even have a further progression*

*towards not natural positions; and the natural inclination of the medium, again, drives back the material at the natural position and so creates a subtle tremor at the ends of the medium. And this tremor is evident by touch and sight.*

"Sonativum" is a medium or abject "able to sound, sonorous". In the first part of the treatise then, we find the description of an oscillation of a medium as the origin of sound. And here we find that Grosseteste considered the sound coming from an oscillating body, or, that the source of a sound was a body having oscillating parts. He probably observed how sounds had produced by using some objects and instruments, and how it was produced by animals and humans; however, he could have analyzed in details some elastic media struck violently. For instance, let us imagine he had observed laterally the oscillations of a thin blade: when the oscillations have a low frequency, we can easily perceive it by eyes, because our eye is able to see clearly the oscillations up to about 20 Hertz. This situation corresponds to a low sound.

It would be great to guess that Grosseteste had observed a tuning fork (diapason) but this instrument was invented in the 1711 by the British musician John Shore [4].

After this connection between sound and vibrations, Grosseteste continues with the sound propagation:

> Hunc tremorem minutarum partium necessario consequitur in egressione a situ naturali extensio earumdem partium secundum diametrum longitudinalem et constrictio secundum diametrum transversalem; et in reversione ad situm naturalem accidit e contrario abbreviatio diametri longitudinalis et majoratio diametri transversalis. Et haec motio sonativi secundum extensionem et contractionem in partibus minutis, quae consequitur motum localem tremoris est sonus vel velocitas naturalis ad sonum. Et cum tremunt partes sonativi movent aerem sibi contiguum ad similitudinem sui motus et pervenit usque ad aerem sibi connaturalem in auribus aedificatum et fit passio corporis non latens animam et fit sensus auditus.

*Such vibrations of each small part of the medium are necessarily a result of their displacement from the natural position, consisting in an elongation of the longitudinal dimension and a contraction of the transversal dimension; and, conversely, when returning towards the natural position, we have a contraction of the longitudinal dimension and an elongation of the transversal one. And this motion of expansion and compression in each part of the medium, where the local motion of vibration is consequent, is the sound or the natural sounding promptness. And when the parts of the sonorous medium are moving, they move the air near them, which having a similar motion, creates a motion which arrives into the ears and this effect on the body is not hidden to the soul and creates the sense of hearing.*

Here, I translated "velocitas naturalis ad sonum", with "natural attitude to sound or sounding promptness". We cannot use here a translation containing a locution such as "speed of the sound". In Latin, "velocitas" means velocity in the sense of promptness [5]. In this Grosseteste's discussion, it is quite interesting the propagation of the sound in the air from a vibrating source. The sound arrives to the air inside the ears and then, it is affecting the body, produce the sense of hearing.

We find also a discussion on "diameters"; I rendered "diametrus", medieval variant of diameter, with "dimension", considering it as a thickness or a width. We can imagine that, instead of a tuning fork, Grosseteste experimented with some wires or metallic strips, bent to form round or elliptic rings, to observe the vibration of a medium prompt to sound when stricken. In the case of such shapes, the vibrations can elongate one of the axes and reduce the other and vice versa.

I preferred to consider "diametrus" as "dimension", in order to have a text closer to a general description of elastic materials. In this case, the phenomenon described by Grosseteste can be similar to the Poisson effect of elastic materials. Let us suppose an elastic material and a bar made of it. When the bar is stretched we see usually that to an extension in the direction of the applied tension, corresponds a contraction in the perpendicular directions. When a material is compressed in one direction, it usually tends to expand in the other two directions perpendicular to the direction of compression. The Poisson's ratio measures this effect. The Poisson's ration is positive in the usual abovementioned behavior of materials [6].

The treatise of Grosseteste continues remarking that the reasons for a medium to sound are two, "either the motive force is internal the very sounding medium or external". In the case that it is internal, it means that it is coming from a voluntary action on breath and articulators of speech. A proper setting of them gives to the voice its appearance and perfection.

After this discussion of the sound and its generation Grosseteste starts with phonetics. In particular he tells that the shape of the letters in the grammar is coming from a representation of some internal settings assumed when pronouncing them. In such a way, the grammar is imitating the Nature. Moreover, the letters of the different languages have symbols which are only accidentally different, not substantially [7, pag.192].

However, if the written letters are in their shapes representing the motions of breath and articulators when pronouncing the sound of them, we need to assimilate the possible motions in some types, each type representing a vocal sound. Then Grosseteste continues his treatise with the list of motions, subdividing them in seven types according to their partial or total similarity.

> Sed motus assimilati sibi in toto et in parte sunt septem: motus rectus, circularis et dilatationis et constrictionis. Haec enim duo non differunt, nisi sicut motus rectus ante et retro motus circularis super centrum motum recte, et motus circularis super centrum motum circulariter; et similiter motus dilatationis et constrictionis super centrum motum recte et super centrum motum circulariter.

*But motions, after being assimilated, totally or partially, are seven: and they are straight motion, circular and of expansion and contraction. Of these, two do not differ except in the direction forwards and backwards of the straight motion, (then we have the) circular motion about a center which is moving straight, and the circular motion about a center in circular motion; and likely, the motion of expansion and contraction over a center on straight motion and over a center in circular motion.*

The assimilated motions are seven, because Grosseteste wants to represent the seven vowels of Greek. Let us remark that the seven Grosseteste's types of motion are not the seven motions (up/down, right/left, forward/back, rotary), proposed by Philo of Alexandria in his treatise On Creation. The Grosseteste's motions are quite complex.

I rendered the Latin text supposing Grosseteste was proposing a combination of motions. The seven motions are as in the following. Three motions are the straight motion, in the two directions, forwards and backwards, and the circular motion about a center at rest. The fourth is the circular motion about a center which is moving on a straight line. This is the description of a cycloid, even prolate or curtate. Let us remember that a cycloid is the curve traced by a point on the rim of a circular wheel as the wheel rolls along a straight line. It is then a curve generated by a curve rolling on another curve. After, the fifth motion given by Grosseteste is a circular motion about a center in circular motion. This seems the description of an epicycloid, which is a plane curve produced by tracing the path of a chosen point of a circle, called an epicycle, which rolls around a fixed circle. This is the motions of the planets in the heavens as described by Ptolemy, well known by scholars such as Grosseteste. The last two are periodic motions, of expansion and contraction, wavelike motions, on a straight line and on a circumference.

These Grosseteste's types of motion seem suitable to be applied to approximately describe, besides the motion of celestial bodies, how parts of some machines can move. As previously told, it is possible that during the renaissance of sciences stimulated by the translation of texts from other cultures, such as the ancient Greek and Arabic works, some scholars started the developments of a scientific language, able to describe what happens in mechanics and technology.

After this description of the seven motions, Grosseteste tells that the ancient Greek set seven vowels according to them. In fact, some of these motions are possible to imagine but difficult or actually impossible to render with the voice. And then he concludes that "just five motions remain, which are possible or operationally feasible". He tells about the motions associated with letters J, V,A, and R, because, as remarked in [8], the art of grammar imitates the nature, and nature does everything in the best way possible, and then the letters of the alphabets have a shape representing the motions of articulators when we are speaking.

For instance, to create A: *Motus vero dilatationis et constrictionis super centrum motum recte motu recto subtendit basim trigoni. Et omnis punctus, cum sit dilatatio, quia movetur, motu duplici, describit unum latus trigoni a basi usque ad conum; et cum sit constrictio, describit reliquum latus a cono usque ad basim; et ita fit figura A.* *And each point, which is moving in such a double motion, when there is the expansion, describes one of the sides of the triangle from the base up to the cone, and when there is the constriction, describes the other remaining side from the cone to the base; and then it is given letter A.* Grosseteste continues with a discussion of the consonants, *quasi cum alio sonans; et quasi per se non possit audiri, cum eius generatio praecedat, vel subsequatur tempore generationem vocalis; consonants, which are so called because they seems to sound with another, and it is not possible to ear by themselves, but by generation of a vowel in the following occurrence.*

About the formation of a syllable, Grosseteste writes:

> Ad hoc respondeo: quod virtus motiva, qua formatur vocalis continue a principio syllabae usque ad finem eius, inclinat spiritus et instrumenta ad formandum sonum vocalis sibi similem et etiam movet spiritus et instrumenta. Cum autem dictam inclinationem concomitatur inclinatio aliqua ad formandum sonum consonantis, egreditur in spiritibus et instrumentis motus unus compositus proveniens a duabus inclinationibus, sicut cum ponderosum inclinatur ad motum deorsum, et cum hoc impellitur ex transverso, consequitur in ipso motus egrediens a diversis inclinationibus aliis a motu naturali. Sed quia continua est inclinatio motus naturalis, semper est reversio ad motum naturalem. Manifestum est igitur, quod in motu, quo formatur sonus consonantis est motus et inclinatio ad formandum sonum vocalis materialis et ita in sono consonantis est sonus vocalis materialiter; est tamen sonus naturalis sicut motum soni consonantis, sicut inclinatio ponderosi naturalis, cum impellitur ponderosum ex transverso, magis, est de motiva inclinatio multociens quam violenta et eadem plus dat motui actuali speciem et formam, quam inclinatio naturalis.

*To this I reply: the motive force, which is giving the vocalization, from the beginning of the syllable to its end, inclines the breathing and the articulators to create the vocal sound like its sound, and therefore moves breathing and articulators consequently. When, however, the said inclination is concomitant to reproduce the sound of a consonant, a composed motion resulting from two inclinations exits from the breathings and movement of the articulators, as it happens when a heavy body tends to move downwards, and it is pushed transversally, and the heavy body moves on a motion following some inclines different from the natural movement. Since, however, the inclination of a natural motion is continuous, the movement is always returning to the natural one. It is clear, therefore, that in the movement, by which the sound of consonant is*

*formed, there is the inclination to form the vowel sound considerably, and so in the sound of a consonant, there is the sound of a vowel substantially; at last, a natural sound is like the motion of the sound of a consonant, like the natural inclination of a heavy body pushed transversally, it is the motive inclination, several times excited, however not vehemently, that gives features and forms to the actual motion, rather than the natural inclination.*

Here Grosseteste uses again the analogy with the motion of a heavy body, which is falling or which is falling after receiving a transversal push. In the first case, we are pronouncing a vowel, the natural motion. When we have a combination of two motions, horizontal and vertical, we have a syllable, where the natural motion is altered by the consonant. Let us note that Grosseteste is also observing that the body returns to the natural falling. Of course, this is rough description of the superposition of vertical and horizontal motions, and of the fact that vertical acceleration prevails. In any case, the initial conditions of motion are giving the "shape" to the motion.
Grosseteste ends his treatise with some further considerations on consonants and semivowels.
The De Generatione Sonorum by Grosseteste is known and studied for the history of phonetics and music. The fact that it contains some interesting discussions on motions is, in my opinion, not properly considered. In this treatise it is clear the Grosseteste widely used the combination of motions, rectilinear and circular, and vertical and transversal. For this reason, a further study of the Grosseteste's works can help in understanding the development of the language of physics by the medieval scholars.